# Noncontrast free-breathing respiratory self-navigated coronary artery cardiovascular magnetic resonance angiography at 3 T using lipid insensitive binomial off-resonant excitation (LIBRE)

Jessica AM Bastiaansen[*,1], Ruud B van Heeswijk[1], Matthias Stuber[1,2], Davide Piccini[1,3]

[1] Department of Diagnostic and Interventional Radiology, Lausanne University Hospital and University of Lausanne, Lausanne, Switzerland

[2] Center for Biomedical Imaging, Lausanne, Switzerland.

[3] Advanced clinical imaging technology, Siemens Healthcare AG, Lausanne, Switzerland.

[*]To whom correspondence should be addressed: Jessica AM Bastiaansen, Department of Radiology, University Hospital Lausanne (CHUV), Rue de Bugnon 46, BH 8.84, 1011 Lausanne, Switzerland, Phone: +41-21-3147516, Email: jbastiaansen.mri@gmail.com

**ABSTRACT**

**Background:** Robust and homogeneous lipid suppression is mandatory for coronary magnetic resonance angiography (MRA) since coronary arteries are commonly embedded in fat. However, effective large volume lipid suppression becomes more challenging when performing radial whole-heart coronary MRA for respiratory self-navigation and the problem may even be exacerbated at increasing magnetic field strengths. Incomplete fat suppression not only hinders a correct visualization of the coronary vessels and generates image artifacts, but may also affect advanced motion correction methods. The aim of this study was to evaluate a recently reported lipid insensitive MRI method when applied to a noncontrast self-navigated coronary MRA acquisitions at 3T, and to compare it to more conventional fat suppression techniques.

**Methods:** Lipid insensitive binomial off resonant excitation (LIBRE) radiofrequency (RF) excitation pulses were included into a self-navigated 3D radial GRE coronary MRA sequence at 3T. LIBRE was compared against a conventional fat saturation (FS) and a binomial 1-180°-1 water excitation (WE) pulse. First, fat suppression of all techniques was numerically characterized using Matlab and experimentally validated in phantoms and in legs of human volunteers. Subsequently, free-breathing self-navigated coronary MRA was performed using the LIBRE pulse as well as FS and WE in ten volunteers. Myocardial, arterial and chest fat signal-to-noise ratios (SNR), as well as coronary vessel conspicuity were quantitatively compared among those scans.

**Results:** The results obtained in the simulations were confirmed by the experimental validations as LIBRE enabled near complete fat suppression for 3D radial imaging in vitro and in vivo. For self-navigated whole-heart coronary MRA at 3T, fat SNR was significantly attenuated using LIBRE compared with conventional FS. LIBRE increased the RCA vessel sharpness significantly ($37 \pm 9\%$ (LIBRE) vs. $29 \pm 8\%$ (FS) and $30 \pm 8\%$ (WE), both $p<0.05$) and led to a significant increase in the measured RCA vessel length to ($83 \pm 31$ mm (LIBRE) vs. $56 \pm 12$ mm (FS) and $59 \pm 27$ (WE) $p<0.05$).

**Conclusions:** Applied to a respiratory self-navigated noncontrast 3D radial whole-heart sequence, LIBRE enables robust large volume fat suppression and significantly improves coronary artery image quality at 3T compared to the use of conventional fat suppression and water excitation.

**BACKGROUND**

Whole-heart coronary MRA is usually performed during free-breathing while the respiratory motion is monitored and gated using a navigator placed on the lung-liver interface, or, more recently, through self-navigation by deriving the respiratory displacement from the acquired data themselves (1–7). To cover the entire coronary arterial tree in a single scan, these acquisitions require a large volumetric coverage, and thus also a large-volume fat suppression.

Large volume fat suppression is of vital importance to generate contrast between the coronary lumen blood pool and epicardial fat, in which the coronary arteries are embedded (8). If fat suppression is suboptimal, the residual lipid signals may hinder the correct anatomical visualization of the coronary vessels and lead to artifacts in the image, while, for self-navigation, it may also degrade the signal quality used for tracking the respiratory displacement and thus degrade the motion correction (4). Firstly, using respiratory motion correction for the heart, motion artifacts in the image will inevitably occur and originate from static structures such as the chest wall. These unwanted signals will unfortunately be amplified if fat signal from the chest is not adequately attenuated. Secondly, and owing to the self-navigation concept where a single profile in k-space informs about respiratory displacement of the anatomy in the entire field of view, the signals from the ventricular blood-pool and that from incompletely suppressed fat may not easily be discriminated. As a result, tracking of the left-ventricular blood pool may become erroneous (4). However, homogeneous and effective fat suppression for such a large volume is quite challenging, mainly due to magnetic field inhomogeneities in the large field of view. This is even more prevalent when moving from 1.5T to 3T and may be amplified for radial whole-heart imaging where each acquired k-space profile traverses through the center of k-space, which represents the average signal of the excitation volume. Therefore, and even if magnetic field inhomogeneities could adequately be accounted for, $T_1$-recovery after a conventional fat saturation pre-pulse may still lead to sub-optimal fat suppression for radial imaging.

At 3T, the use of water selective radiofrequency (RF) excitation pulses may solve the problem related to the $T_1$ recovery of lipid signals (9–12), especially for radial sampling schemes. However, the improved fat suppression capabilities may come at the expense of increased RF pulse durations, since typically, the fat suppression bandwidth of conventional binomial water excitation (WE) pulses increases with the number of sub-pulses. To address this limitation, a non spatially selective water excitation pulse was proposed that demonstrated robust fat suppression at 3T with total RF pulse durations as short as 1.4 ms (13). More recently these pulses were further shortened to a total duration of 1.0 ms (14).

The aim of the current study was to implement and exploit lipid-insensitive binomial off-resonance excitation (LIBRE) pulses for fat suppression in 3D radial noncontrast self-navigated coronary MRA at 3T, and to compare the results to those obtained with conventional fat suppression methods. To this end, a numerical comparison was made between a conventional on-resonance binomial 1-180°-1 water excitation (WE) pulse, a chemical shift selective fat saturation pulse (FS), and the off-resonance LIBRE pulse (Fig. 1a-c). A quantitative comparison between these three different fat suppression schemes was then made using 3D radial acquisitions in phantoms and knees of healthy volunteers, as well as whole-heart respiratory self-navigated coronary MRA in healthy volunteers.

**METHODS**

*Theory*

The LIBRE pulse used in this study (13) consists of a pair of low power rectangular pulses, each having the same RF excitation angle α, RF excitation frequency $f_{RF}$, and sub-pulse duration τ. In the absence of $B_0$ field inhomogeneities, optimal fat suppression is predicted when the following condition is met:

$$\tau = \sqrt{1-(\alpha/2\pi)^2}/(f_{RF}-f_{fat}) \qquad (Eq.\ 5)$$

To illustrate this condition, assuming a fat frequency $f_{fat}$=-440 Hz at 3T, and in the absence of field inhomogeneities, the LIBRE RF excitation frequency $f_{RF}$ (that leads to optimal fat suppression) was plotted as function of the sub-pulse duration for an RF excitation angle of 18° and 90° (Fig. 2). Field inhomogeneities may broaden the line width or the position of the fat resonance frequency, therefore a range of $f_{fat}$ from -400Hz to -480 Hz was also evaluated.

*Numerical simulations*

To predict and calculate the magnetization behavior of cardiac tissue using the three different fat suppression methods in the presence of $B_1$ and $B_0$ inhomogeneities, numerical simulations were performed in Matlab (The MathWorks, Inc., Natick, MA, United States). The water and fat magnetization components were evaluated as function of tissue frequency $f_{tissue}$, magnetic field inhomogeneities $\Delta B_0$, and RF excitation angle α. Simulations over a range of RF angles also instruct about $B_1$ inhomogeneities. The numerical simulations, similar to those described before (13), were extended to take the relaxation times $T_1$ and $T_2$ into account, as well as repeated excitations, using the Bloch equations. $T_1$ of myocardial blood was assumed to be 1932 ms, and $T_2$ was set to 275 ms (15). A repetition time (TR) of 5.1 ms was used reflecting the TR in the MRA protocol.

Because the acquisition window for coronary imaging in healthy volunteers in previous studies (4,6) was typically on the order of 100 to 120 ms depending on the duration of the mid-diastolic resting phase of the heart, the number of simulated RF excitations was set to 24. The simulations were performed for RF excitation angles ranging from 0° to 50°, to simulate a range of typical RF excitation angles of a GRE acquisition. Simulated tissue frequencies $f_{tissue}$ were ranging from -600 Hz to 600 Hz to adequately encompass frequencies of both water and fat. A LIBRE pulse with RF sub-pulse duration τ of 1.1 ms was chosen because these pulse properties result in a similar TR compared with a conventional WE. The transverse magnetization was set to zero after each excitation to mimic perfect spoiling. Plots were made to visualize the transverse magnetization as function of the RF excitation angle and different tissue frequencies $f_{RF}$. To compare the magnetization behavior following repeated LIBRE excitations, numerical simulations with identical parameters and parameter ranges were also carried out using a conventional 1-180°-1 water excitation (WE) pulse (16), and a frequency selective pulse for fat saturation (FS) (17). The WE pulse consisted of two on-resonance rectangular sub-pulses, each with duration of 0.5 ms and separated by 1.1 ms to allow for a 180° phase evolution between water and fat. The FS simulation was performed by assuming a Gaussian-shaped RF pulse with duration of 5.12 ms, with RF offset frequency of -407 Hz and RF excitation angle of 110°. The exact structure of this Gaussian-shaped pulse was obtained from the sequence product source code. In the simulation for FS, the 24 on-resonance RF excitation pulses during the acquisition were set to a duration of 0.3 ms, as in the product sequence. The choices for FS and WE pulse parameters matched those used in experiments. The fat suppression bandwidth was defined as 10% of the maximum transverse magnetization.

*In vitro exams*

The LIBRE RF pulse implementation was integrated into a pre-existing prototype 3D radial spoiled GRE sequence adapted for self-navigated free-breathing coronary MRI. The radial trajectory follows a 3D spiral phyllotaxis pattern as described mathematically in (18) with a golden angle rotation about the z-axis. Each 3D spiral segment was composed of 24 radial k-space lines to match the simulations. In phantoms, experiments were performed with 1) a LIBRE pulse with sub-pulse duration ($\tau$) of 1.1 ms and an RF frequency offset ranging from 300 Hz to 700 Hz in steps of 20 Hz, 2) a conventional WE pulse with a binomial 1-180°-1 pulse pattern (16), and 3) a conventional FS method that uses a CHESS (17) pulse to null the fat signal prior to the imaging sequence. Data volumes from phantoms were acquired on a clinical 3T MRI system (MAGNETOM Prisma$^{FIT}$, Siemens Healthcare, Erlangen, Germany). 3D volumes with an isotropic voxel size of 1.1 mm$^3$ were acquired with a field-of-view (FOV) of 220 x 220 x 220 mm$^3$, matrix size 192$^3$, RF excitation angle = 18°, TE/TR(FS)=1.6/3.2 ms, TE/TR(LIBRE)=2.5/5.1 ms, TE/TR(WE)=2.3/4.8 ms and a 40 ms adiabatic T$_2$-preparation (19), and using a 15-channel Tx/Rx knee coil. To achieve a 20% sampling of the Nyquist criterion for 3D radial acquisitions as recommended in (18), a total of ~12k lines were acquired. The cylindrical phantom consisted of three compartments containing mixed solutions of agar, NiCl$_2$ (Sigma Aldrich, St. Louis, MO), and baby oil (Johnson and Johnson, New Brunswick, NJ), in order to mimic the magnetic relaxation properties (T$_1$, T$_2$) of muscle, blood, and fat.

Table 1. MR sequence details and parameters

| Parameter | FS | WE (1-180°-1) | LIBRE[a] |
|---|---|---|---|
| RF duration (total in ms) | 0.50 | 1.7 | 2.2[a] |
| TE (ms) | 1.6 | 2.3 | 2.5 |
| TR (ms) | 3.2 | 4.8 | 5.1 |
| RF Pulse offset (Hz) | 0 | 0 | 480 |
| RF excitation angle (°) | 18 | 18 | 18 |
| T2 preparation duration (ms) | 40 | 40 | 40 |
| Matrix size | $192^3$ | $192^3$ | $192^3$ |
| Field of view | $(220\ mm)^3$ | $(220\ mm)^3$ | $(220\ mm)^3$ |

[a] The LIBRE pulse has a flexible duration and shorter variations are possible as indicated in (13).

*In vivo knee exams*

To validate the in vitro results and to separate the confounding effects that motion may have on the in vivo results, the fat suppressing performance of LIBRE combined with 3D radial was ascertained in static muscle and fat tissue in vivo. Therefore, the same experiments as those performed in the phantoms were repeated in knees of human volunteers (n=3) by setting the LIBRE $f_{RF}$ to the optimal frequency derived from the phantom experiments, i.e. 480 Hz for a sub-pulse duration ($\tau$) of 1.1ms (Fig. 1d). All volunteers provided written informed consent and local ethical authorities approved this study.

*In vivo free-breathing whole-heart coronary MRA*

Noncontrast whole-heart coronary MRA was performed in 10 healthy adult volunteers on the same clinical 3T MRI system using the LIBRE, FS and WE protocols in randomized order (Fig. 1). For this purpose, a 3D radial imaging sequence as described above was used for coronary artery imaging by enabling respiratory-self-navigation and ECG triggering (4,6,20,21). Whole-heart volumes were acquired during free-breathing. To achieve a 20% sampling of the Nyquist criterion for 3D radial acquisitions as recommended in (18), a total of ~12k radial profiles were acquired in a segmented fashion per 3D scan while 20-28 profiles were collected per heartbeat. The amount of profiles acquired per heartbeat varied across volunteers. In each volunteer we determined the duration of the mid-diastolic cardiac resting phase by visual inspection of a midventricular 2D cine scan. Then it was determined how many segments could be acquired during this period using the protocol with the longest TR, which was LIBRE with a TR of 5.1ms. Subsequently all protocols were performed using the same number of segments per heartbeat and volunteer. The heart rate and average acquisition windows were recorded in each volunteer. The whole-heart coronary MRA acquisitions were respiratory motion-corrected and reconstructed directly at the scanner using a superior-inferior (SI) projection acquired at the beginning of every segment (every heartbeat) as previously described (4). The reconstruction was based on 3D gridding and the reconstruction time was below 1 min.

*Data analysis*

All MRI datasets were directly reconstructed at the scanner using the sum-of-squares of all channels and the gridding algorithm provided by the vendor. In the phantom experiments, the signal-to-noise-ratio (SNR) was calculated in compartments containing fat to evaluate the level of fat suppression, the noise was measured in regions containing air, and the contrast-to-noise ratio (CNR) between compartments mimicking myocardium and blood. In volunteer experiments performed in the leg, the SNR was calculated in compartments containing muscle tissue and fat for comparison among the three different fat suppression methods. In whole-heart coronary MRA, the SNR was calculated in ROIs drawn on the myocardium at the level of the interventricular septum, in the chest fat, and in the left ventricular blood pool. Noise was calculated in regions containing air, outside the volunteers within the field of view. The CNRs were computed to evaluate the level of contrast of the blood pool relative to the myocardium, as well as blood versus fat. All images were analyzed in ImageJ (National Institutes of Health, Bethesda, Maryland, USA). Coronary vessel sharpness and vessel length were quantified (22) in both the left anterior descending (LAD) artery and the right coronary artery (RCA) for all acquired whole-heart volumes. Vessel sharpness was computed for the same length for all methods (proximal 4cm). Coronary reformats were also generated to visualize and compare the structure of the LADs and RCAs in all volunteers and across techniques. A paired Student's t-test, corrected for multiple comparisons, was performed on phantom and volunteer data and $p<0.05$ was considered statistically significant. All data are represented as average ± one standard deviation.

**RESULTS**

*Theory and numerical simulations*

Evaluation of Eq. 1 showed a range of parameter combinations that indicate optimal fat suppression (Fig. 2). Assuming a fat resonance frequency of -440 Hz, and in the absence of field inhomogeneities that broaden the line shape of the fat resonance, the combination of a sub-pulse duration of 1.1 ms and an RF excitation frequency of 469 Hz provides complete fat nulling. Assuming fat resonances at -400 Hz or -480 Hz, the optimal $f_{RF}$ changes to 507 Hz and to 429 Hz respectively (Fig. 2). Although Eq. 1 shows that the RF excitation angle affects the optimal combination between the LIBRE frequency and duration, an increase in the RF excitation angle to 90° had a minor influence on the choice of optimal LIBRE parameter combinations (Fig. 2).

Numerical simulations demonstrated that at an RF excitation angle of 18° the transverse magnetization of on-resonance water is highest, with a fat suppression bandwidth on the order of 238 Hz ($f_{RF}$=479 Hz, $\tau$=1.1 ms), compared to 32 Hz and 128 Hz using WE and FS respectively (Fig. 3). The magnitude of the transverse magnetization observed at a tissue frequency of ~0 Hz demonstrated that a similar range of RF excitation angles may be used across techniques to achieve the same excitation behavior. The change in transverse magnetization over a range of RF excitation angles can also be interpreted as a measure of the sensitivity of the investigated sequences to $B_1$ inhomogeneities.

*Experimental results in phantoms and legs of human volunteers*

The phantom experiments showed a clear difference across different fat suppression techniques (Fig. 4a). WE performed well at the center of the cylindrical phantom, but its fat suppression efficiency degrades moving towards the phantom boundaries, i.e. regions that suffer from magnetic field inhomogeneities. FS performed poorly in this 3D radial acquisition, with signal leaking across the boundaries of the air-phantom interface. LIBRE suppressed fat homogeneously in the entire phantom, including the boundaries. A frequency calibration of the LIBRE pulse in a phantom demonstrated that a range of $f_{RF}$ from ~400 Hz to ~500 Hz resulted in optimum fat suppression (Fig. 4b). The lowest SNR of fat (8.3 ± 0.9) was obtained using LIBRE with an $f_{RF}$ of 460 Hz, compared with FS (48.1 ± 3.4) and WE (31.8 ± 2.1) (Fig. 4b), both p<0.05. The CNR between the inner and outer compartments of the phantom, that mimic blood and myocardial tissue respectively, was not significantly different across the different techniques (Fig. 4c), (p=NS).

Measurements in the leg showed a similar behavior as in the phantom experiments (Fig. 4d) with the fat being homogenously suppressed using LIBRE. The SNR of fat was significantly decreased using LIBRE (9.9 ± 2.2) compared with FS (26.6 ± 6.9, p<0.005) and WE (25.1 ± 6.1, p<0.005) (Fig. 4e). The SNR of skeletal muscle tissue was similar comparing LIBRE and WE, but decreased using FS.

*In vivo free-breathing whole-heart coronary MRA*

Free-breathing coronary MRA was successfully performed in all volunteers without complications with an average scan time of 8.6 ± 1.5 min. Volunteers had an average heart rate of 64 ± 8 BPM. The average data acquisition window was 81 ± 17 ms (FS), 122 ± 26 ms (WE) and 129 ± 28 ms (LIBRE). The left and right coronary systems could be visualized clearly in all cases, but not with all techniques. A clear difference across the different techniques can be observed in the coronary reformats where both the RCA and LAD are visualized (Fig. 5). Compared to the use of WE and FS, coronary

LIBRE leads to an improved visualization of the coronary arteries. In addition, large volume fat suppression can be achieved using LIBRE, as can be seen from the decrease in signal from fatty tissue in the back and chest of a volunteer (Fig. 6, red and green arrows). The three different fat suppression methods also had an effect on the signal behavior in the SI projections (Fig. 6, bottom row).

The SNR of the myocardium was similar across all techniques, with no statistically significant differences found (p=NS). The blood SNR was 86 ± 35 using LIBRE and was significantly decreased to 51±18 using FS (p=0.005) and to 55±12 using WE (p=0.01) (Fig. 7a). Fat SNR was significantly increased from 14±8 to 46±18 using FS (p=0.01) and to 19±7 using WE (p=NS) (Fig. 7a). CNR between blood and myocardial tissue significantly decreased from 33±12 using LIBRE to 16±17 using FS (p=0.01) and to 16±3 using WE (p=0.01) (Fig. 7b). The CNR between blood and fat tissue was significantly increased from 24±12 using FS to 76±41 using LIBRE (p=0.01) (Fig. 7b). The vessel sharpness of the RCA and LAD was significantly improved (Fig. 7c, $p<0.05$ in all comparisons) using LIBRE (37 ± 9% and 34 ± 7%, respectively), compared with FS (29 ± 8% and 24 ± 6%, respectively), and with WE (30 ± 8% and 27 ± 7%, respectively). In addition, the measured vessel length of both the RCA and LAD was significantly increased (Fig. 7d, $p<0.05$ in all comparisons) using LIBRE (83 ± 31 mm and 98 ± 35 mm), compared with FS (56 ± 12 mm and 54 ± 25 mm), and with WE (59 ± 27 mm and 61 ± 21 mm).

## DISCUSSION

In this study we implemented a previously published LIBRE pulse (13) in combination with a respiratory self-navigated 3D radial imaging sequence (4) and demonstrated its effectiveness for fat suppression and its application for coronary MRA at 3T in volunteers without the use of contrast agents.

The LIBRE pulse was optimized in phantoms and legs and experiments showed that improvements in fat suppression for radial imaging were consistent with those found for a Cartesian approach as originally reported (13). Radial imaging is inherently more sensitive to incomplete fat suppression, especially at higher magnetic field strengths, where field inhomogeneities are typically accentuated. The increased fat suppression bandwidth of LIBRE led to a near-complete nulling of the fat signals in large 3D volumes and outperformed conventional fat suppression methods such as CHESS based spectral fat saturation (17) and binomial 1-180°-1 water excitation (16).

When assessing the coronary MRA results obtained using different fat suppression techniques, it can be appreciated that LIBRE improved the quality of the vessels conspicuity. A quantitative comparison revealed significant improvements when using LIBRE in terms of vessel sharpness, detectable vessel length, SNR and CNR. The underlying reason for differences in final image quality across three techniques is most likely a combination of two factors that cannot be decoupled experimentally. The first one relates directly to the significantly improved fat suppression using the LIBRE method as measured also in the static phantom and volunteer scans. Secondly, this improved fat suppression may have benefitted the motion tracking and motion correction used in respiratory self-navigation. As fat is more homogeneously and more completely suppressed, the SI projections used for respiratory motion tracking contain less residual fat signal from static structures such as the chest wall and the arms that may hinder a reliable motion detection. In addition, the artifacts due to motion correction performed on residual static fat signal from the chest may be less pronounced.

Although 3T coronary MRA has also been performed using SSFP in some studies (23,24) the most frequently used acquisition method currently remains GRE with contrast agent injection (25–28). Other methods such as Dixon or IDEAL have been presented as an alternative and would ideally apply with radial imaging, but may require the acquisition of multiple images at several echo times to adequately separate water and fat images (29–31). The presented LIBRE method requires the acquisition of a single image. However, it was recently demonstrated that a two-point Dixon implementation of a GRE sequence led to an improved coronary MRA image quality over conventional fat suppression without increasing scan time (32). This acquisition was also performed without contrast agent injection and may provide a promising alternative to our proposed technique. However, a comparison with water-fat separation techniques was not performed in the current study.

The LIBRE pulse is a spatially non-selective water excitation pulse with a broad fat suppression bandwidth. These properties render the method highly suitable for the acquisition of large 3D volumes. Cardiac MRI examinations are typically complex, because of the anatomical structure, size and spatial orientation of the heart and therefore, MRI acquisitions that cover the entire organ may be preferred over 2D or 3D targeted acquisitions as operator dependency can be minimized and arbitrary views and orientations reconstructed retrospectively.

The fact that the LIBRE pulse is spatially non-selective may pose a limitation for some applications. However, this may also be an advantage when combined with acquisitions that do not require time-consuming oversampling in the phase encoding direction, as is typically the case in 3D radial imaging. As for the self-navigation module, respiratory motion tracking relies on a correct delineation of the blood pool in the heart, which is based on the SI projections. As was observed in this study, a correct delineation of the blood pool may have been hindered using the WE and FS sequences as signal from the fat tissue located in the chest and the back was not always sufficiently suppressed (Fig. 6) at 3T. Not only may unsuppressed fat signal in static structures of the body pose a problem for tracking, it can also affect the final image quality when static tissue with very high signal intensity is incorrectly motion-corrected. Therefore it may be argued that LIBRE did not only improve the homogenous suppression of fat signal, but also improved the tracking of the blood pool, and the motion-correction performed in respiratory-self-navigation. However, using the non-spatially selective LIBRE pulse and the large excitation volume, signals from arms may also be included in the SI projections and affect the accuracy of respiratory motion estimation and correction.

The use of water excitation pulses for radial imaging at 3T may be preferred over fat signal saturation because radial imaging is extremely sensitive to fat signal recovery as each acquired k-space line goes through the k-space center and T1 recovery of fat during the acquisition window will be unavoidable. Therefore, methods that rely on fat signal suppression by applying a frequency selective saturation pulse, or that rely on T1-based nulling of the fat signal following a frequency selective inversion pulse, were shown to work well in Cartesian based imaging approaches (33,34), but may still yield unwanted residual fat signal in the final images using radial imaging. A further optimization of these suppression techniques for this particular 3D radial application in terms of timing, RF excitation angles, and other parameters may have to be performed, but was outside the scope of the current paper. At 3T, the T1 of fat is longer and fat signal suppression using frequency selective inversion pulses or saturation pulses may become more effective, particularly for very short acquisition windows. However, as our acquisition window is still > 80ms, as every profile of a radial sequence goes through the center of k-space, and provided that B0 and B1 inhomogeneities are enhanced at 3T, the benefit of the longer T1 of fat is expected to be minimal. An improved performance of frequency selective inversion pulses or saturation pulses could also be achieved by decreasing the amount of radial lines that are acquired per heartbeat, however this comes at the expense of an increase in scan time. Water excitation can be performed using binomial pulse patterns such as 1-1, 1-2-1, and even 1-3-3-1. An increase in the number of sub-pulses leads to a broadening of the suppression bandwidth, but at the expense of RF pulse duration. Such water excitation pulses may vary in total RF duration from 1.7 to 3.9 ms at 3T, and may thus increase the TR and scan time significantly. LIBRE demonstrated a large fat suppression bandwidth with total RF pulse durations as short as 1.4 ms (13), and current investigations suggest that this may be further decreased to a total RF pulse duration of 1 ms ($\tau$=0.5ms) only (14). Since the optimal duration of the LIBRE pulse is a function of its RF offset and the resonance frequency of the fat (Eq. 1), similar fat suppression may also be achieved at higher magnetic fields strengths with a shorter RF pulse duration. Therefore this LIBRE pulse may be suitable for coronary MRA at field strengths beyond 3T as well (33,35).

Finally, the proposed LIBRE method in combination with free-breathing radical coronary MRA offers a flexible pulse duration while maintaining fat suppression capabilities (13).

Since coronary artery disease (CAD) is a leading cause of death in the developed world, there is a need for a noninvasive imaging method that can detect CAD without using ionizing radiation without any type of contrast media. Moreover an imaging method that can be used as a radiation-free alternative to detect anomalies in the coronary anatomy in children and young adults is highly desirable. A recent study in a pediatric cohort compared the use of respiratory self-navigated coronary MRA with computed tomography (CT) angiography at 1.5T and demonstrated that coronary artery anomalies could be detected on MRA with high accuracy (36). Other patient studies using respiratory self-navigation in a clinical environment have been recently published with promising results using a 1.5T scanner (20,21). LIBRE has the potential of enabling a similar if not superior scan quality also at 3T and, currently, an ongoing clinical study is being performed in heart transplant patients using this technique (37). Future direct comparisons between the LIBRE self-navigated sequence and a gold standard CT scan are of course warranted.

## CONCLUSION

LIBRE water excitation pulses were implemented as part of a 3D coronary MRA radial imaging sequence at 3T and demonstrated an effective and robust fat signal suppression both in vitro and in vivo. The LIBRE method significantly improved coronary artery image quality compared with more conventional fat suppression methods when self-navigated free-breathing noncontrast MRA was performed without the use of a contrast agent.

## LIST OF ABBREVIATIONS

RF, radio frequency; MRI, magnetic resonance imaging; FS, fat suppression; WE, water excitation; LIBRE, lipid insensitive binomial off-resonant excitation; MRA, magnetic resonance angiography;

## DECLARATIONS

*Ethics approval and consent to participate*

This study was approved by local authorities. All volunteers provided written informed consent for participation in this study.

*Consent for publication*

All volunteers provided consent for publication of this study.

*Availability of data and material*

The datasets used and/or analyzed during the current study are available from the corresponding author on reasonable request.

*Funding*

R'Equip SNF grant 326030_150828, and the MagnetoTeranostics project that was scientifically evaluated by the Swiss National Science Foundation (SNSF), financed by the Swiss Confederation and funded by Nano-Tera.ch (project No 530 627). JB received funding from the SNSF (grant number PZ00P3_167871), the Emma Muschamp foundation, and the Swiss Heart foundation.

*Authors' contributions*

JB developed and implemented the LIBRE RF pulses, analyzed the data, and wrote the main draft of the manuscript. JB and DP designed the study and performed the experiments. DP contributed the source code for the respiratory-self-navigated MRI sequence. RH contributed the source code of the $T_2$ preparation module. All authors read, revised, and approved the final manuscript.

*Acknowledgements*

Not applicable

*Competing interests*

DP is an employee of Siemens Healthcare. JB, RH, MS have no competing interests.

**FIGURE 1**

An ECG-triggered 3D radial sequence was used for coronary MRA in healthy volunteers with three different fat suppression methods. a) Conventional fat saturation (FS) using a spectral prepulse to saturate the fat, b) a conventional binomial 1-180°-1 water excitation (WE) pulse, c) a LIBRE pulse. All acquisitions were preceded by a $T_2$-preparation module of 30 ms. The first acquired k-space profile in each ECG-triggered segmented acquisition was Fourier transformed to obtain a superior-inferior (SI) projection image of the thorax that was used for respiratory-self-navigation. The same sequences were also carried out in phantoms and legs of volunteers by disabling the ECG trigger. d) A diagram illustrating the LIBRE pulse timing.

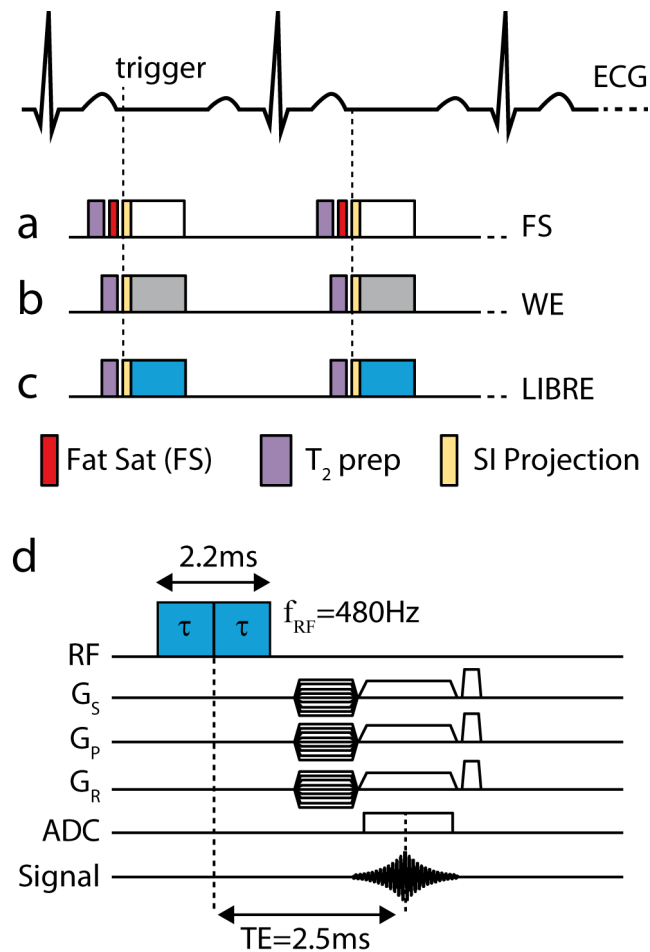

**FIGURE 2**

The relation (see Eq. 1) between sub-pulse duration (τ) and excitation frequency of the LIBRE pulse for optimal fat nulling was plotted using a RF excitation angle of 10° (blue lines) and 90° (dashed red line). Shades of blue indicate a range of fat resonance frequencies, from -400 Hz to -480 Hz. Note that the optimal LIBRE frequency (dashed blue lines) for fat suppression varies according to the resonance frequency of fat and thus illustrates the need for a large fat suppression bandwidth. Similar regions were indicated in the phantom experiments where f$_{RF}$ was varied (Fig. 1b).

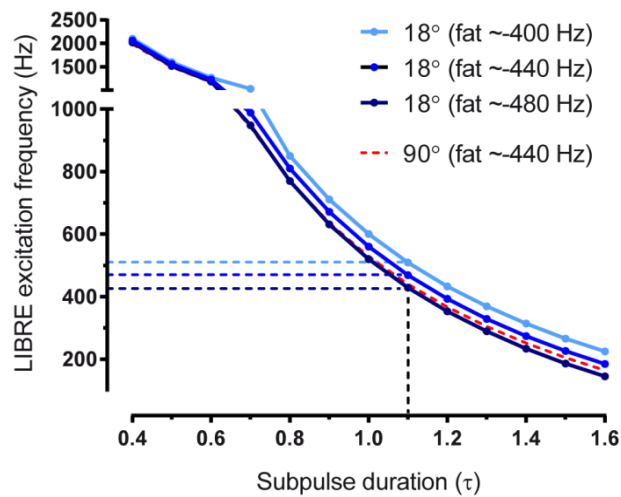

# FIGURE 3

Numerical simulation of the transverse magnetization ($M_{xy}$) as function of RF excitation angle and tissue frequency using three different sequences, a) FS, b) WE, and c) LIBRE. The white arrows indicate the bandwidth of fat suppression (around -440Hz) at a RF excitation angle of 18°. The white and black dashed isolines indicate 10% and 90% of the maximum observed $M_{xy}$, respectively. White arrows indicate the fat suppression bandwidth (up to 10% of maximum $M_{xy}$) that corresponded to 238 Hz (LIBRE), 128 Hz (FS), and 32 Hz (WE).

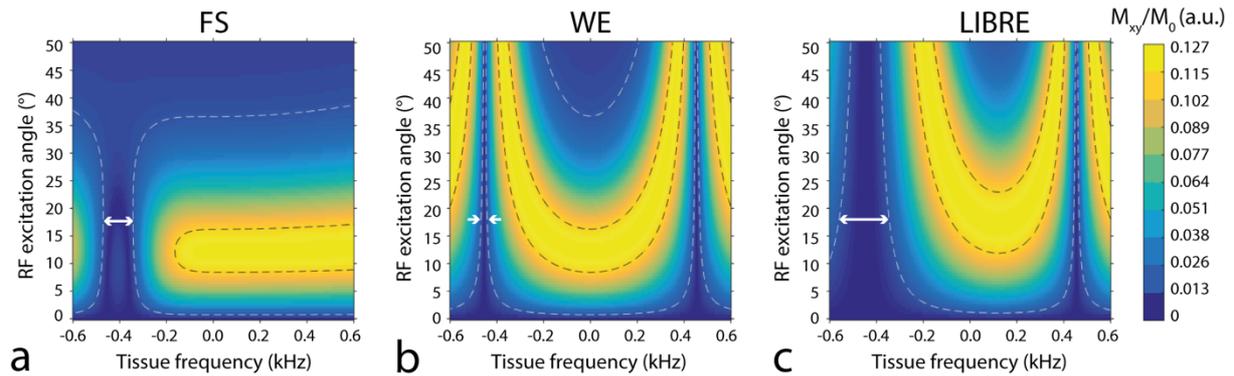

## FIGURE 4

Three different fat suppression methods (Fig. 1) for radial imaging were investigated in phantoms and legs of volunteers. a) Phantom images indicate a clear difference across different techniques. b) Quantitative results obtained after image analysis of the phantom experiments show fat SNR as function of the LIBRE excitation frequency $f_{RF}$ with a total pulse duration of 2.2 ms (blue). The SNR of fat in identical acquisitions using FS or WE are also indicated as a reference. Dashed blue lines correspond to the theoretical optimum for $f_{RF}$ for fat resonating at -400 Hz, -440 Hz and -480 Hz (Fig. 2, Eq. 1). c) The CNR determined between the inner and outer compartment of the phantom that mimic blood and myocardial tissue respectively is comparable across different techniques. d) Leg images obtained in healthy volunteers using three different fat suppression methods. e) SNR of fat and muscle tissue in the leg show a significant decrease of fat SNR using LIBRE, while the SNR of muscle tissue remains similar to WE. Note that the perceived differences between the tissue signals (fat and muscle) in the images and the quantitative SNR plots (c versus d) can be mainly attributed to an increase in the noise in FS. Image intensities were scaled for identical window and level settings. **$p<0.005$

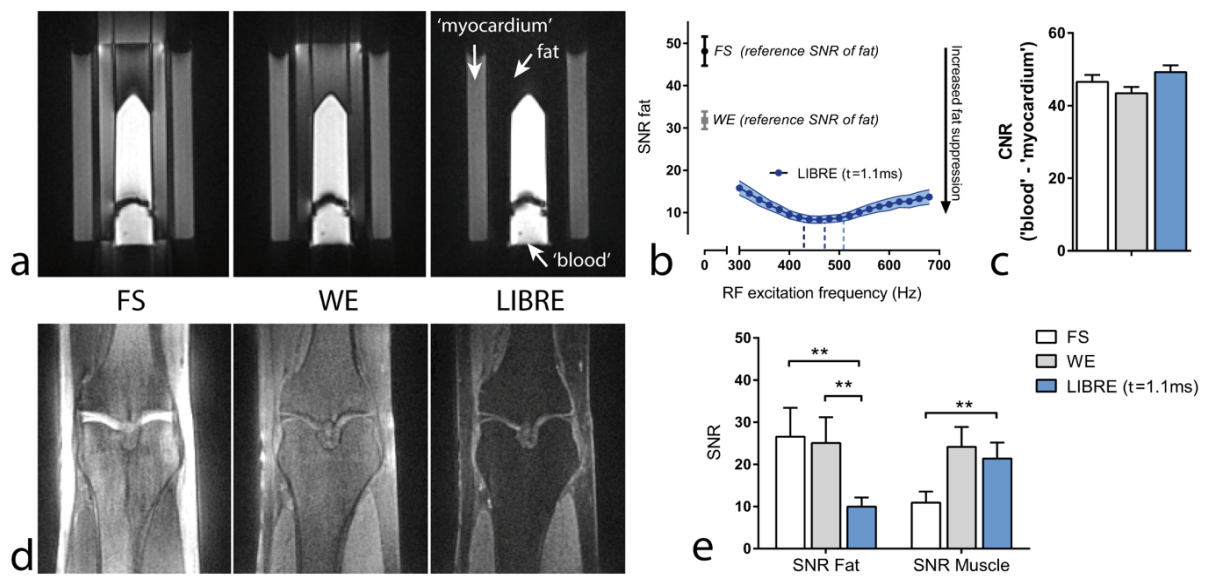

# FIGURE 5

Noncontrast free-breathing coronary MRA was performed at 3T using three lipid nulling methods in healthy volunteers. MR angiograms show the left and right coronary artery system depicting the RCA and the LAD in several volunteers. Using the LIBRE pulse the visualization of the RCA and LAD was improved (yellow arrow), as well as fat suppression (orange arrows) compared with FS and WE. Vessel sharpness as well as vessel length were significantly increased using LIBRE. Window and level are identical in images acquired in each volunteer.

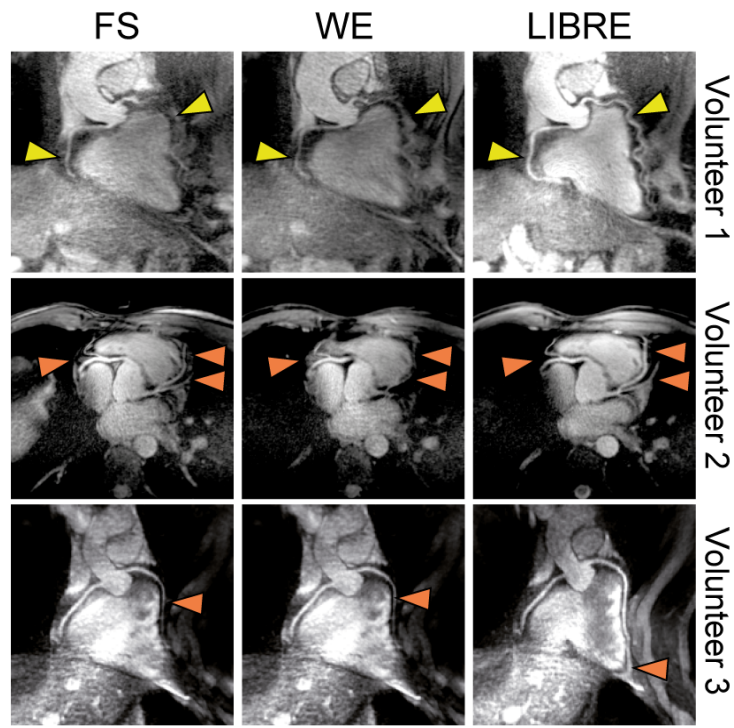

**FIGURE 6**

Sagittal images illustrate the large volume fat suppression in a volunteer using three different methods of lipid nulling (top row) and the corresponding SI projections (bottom row). Note the inhomogeneous fat suppression in the chest (green arrows) and the back (red arrows) using FS and WE compared with LIBRE. Unsuppressed bright fat signal from the chest and the back contribute to the SI projection (bottom row) and may hinder respiratory motion tracking (white lines on SI projections) that relies on a correct delineation of the blood pool in the heart.

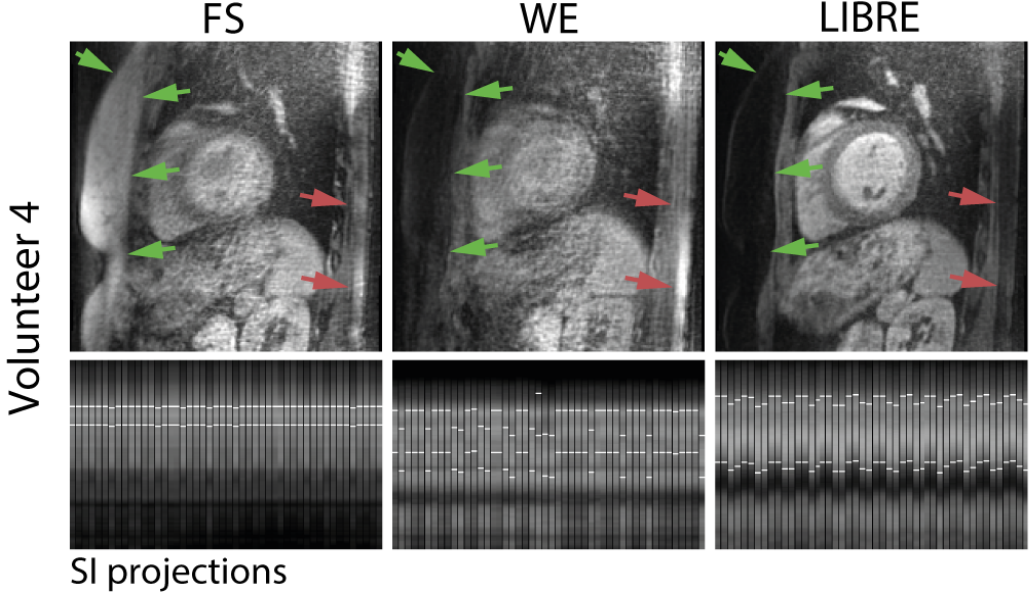

# FIGURE 7

Quantitative endpoints comparing three lipid suppression techniques for whole-heart free-breathing coronary MRA. SNR (a) and CNR (b). Vessel sharpness (c) and vessel length (d) of both the RCA and LAD.

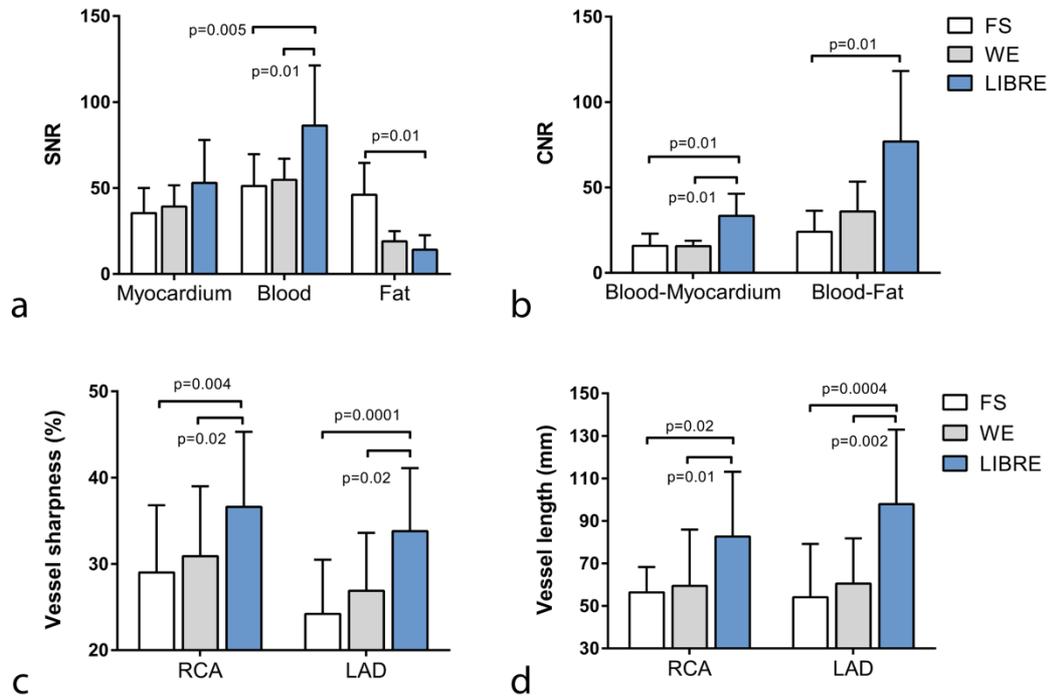